\RequirePackage{etoolbox}
\csdef{input@path}{{style/}{graphics/}}
\documentclass[MSNbibl,nameyear,dvips]{arxstspdf}
\usepackage{multirow}
\usepackage{slashbox}
\usepackage{graphicx}
\usepackage{url,breakurl}

\usepackage{flushend}
\usepackage{stfloats}

\volume{30}
\issue{1}
\pubyear{2015}
\firstpage{26}
\lastpage{39}
\doi{10.1214/14-STS469} 

\makeatletter
\renewcommand{\citep}[1]{(\citeauthor{#1}, \citeyear{#1})}
\makeatother

\begin{document}
\begin{frontmatter}

\title{Le Her and Other Problems in Probability Discussed by Bernoulli,
Montmort and Waldegrave}
\runtitle{Problems in Probability by Bernoulli, Montmort \& Waldegrave}

\begin{aug}
\author[A]{\fnms{David R.}~\snm{Bellhouse}\corref{}\ead[label=e1]{bellhouse@stats.uwo.ca}}
\and
\author[B]{\fnms{Nicolas}~\snm{Fillion}\ead[label=e2]{nfillion@sfu.ca}}
\runauthor{D. R. Bellhouse and N. Fillion}

\affiliation{Western University and Simon Fraser University}

\address[A]{David R. Bellhouse is Professor,
Department of Statistical and Actuarial Sciences,
University of Western Ontario,
London, Ontario N6A 5B7, Canada \printead{e1}.}
\address[B]{Nicolas Fillion is Assistant Professor,
Department of Philosophy,
Simon Fraser University,
Burnaby, British Columbia V5A 1S6, Canada \printead{e2}.}
\end{aug}

%
\begin{abstract}
Part V of the second edition of Pierre R\'emond de Montmort's \emph
{Essay d'analyse sur les jeux de hazard} published in 1713 contains
correspondence on probability problems between Montmort and Nicolaus
Bernoulli. This correspondence begins in 1710. The last published
letter, dated November 15, 1713, is from Montmort to Nicolaus
Bernoulli. There is some discussion of the strategy of play in the card
game Le Her and a bit of news that Montmort's friend Waldegrave in
Paris was going to take care of the printing of the book. From earlier
correspondence between Bernoulli and Montmort, it is apparent that
Waldegrave had also analyzed Le Her and had come up with a mixed
strategy as a solution. He had also suggested working on the ``problem
of the pool,'' or what is often called Waldegrave's problem. The
Universit\"atsbibliothek Basel contains an additional forty-two letters
between Bernoulli and Montmort written after 1713, as well as two
letters between Bernoulli and Waldegrave. The letters are all in
French, and here we provide translations of key passages. The trio
continued to discuss probability problems, particularly Le Her which
was still under discussion when the \emph{Essay d'analyse} went to
print. We describe the probability content of this body of
correspondence and put it in its historical context. We also provide a
proper identification of Waldegrave based on manuscripts in the
Archives nationales de France in Paris.
\end{abstract}

%
\end{frontmatter}

\section{Introduction}\label{sec1}

The earliest extant correspondence between Pierre R\'emond de Montmort
and a member of the Bernoulli family is a letter from Montmort to
Johann Bernoulli dated February 27, 1703, concerning a paper on
calculus that the latter had written for the Acad\'emie royale des
sciences in Paris \citep{e05}. They corresponded sporadically
over the next few years. On April 29, 1709, Montmort sent Bernoulli a
copy of his book on probability, \emph{Essay d'analyse sur les jeux de
hazard}, that he recently had published \citep{e17}. The book is
the first in a series of books in probability published by several
others over the years 1708 to 1718 in what Hald [(\citeyear{e12}), page 191] calls
the ``Great Leap Forward'' in probability. Bernoulli replied with a
gift of a copy of his nephew's doctoral dissertation \citep{e06},
the second book in Hald's ``Great Leap Forward''; Nicolaus Bernoulli's
book dealt with applications of probability. Once Johann Bernoulli
received his copy of \emph{Essay d'analyse}, he sent, on March 17,
1710, a detailed set of comments on the book. In the letter Bernoulli
included another set of comments on \emph{Essay d'analyse}, this one by
his nephew Nicolaus (\cite{e18}, pages 283--303). Thus began a
series of correspondence between Montmort and Nicolaus Bernoulli on
problems in probability. Montmort included much of this correspondence
in Part V of the second edition of \emph{Essay d'analyse} \citep{e18}. The correspondence between Montmort and Nicolaus Bernoulli after
1713, left unpublished and largely ignored by historians, contains
scientific news and further discussion of problems in probability. The
major topic is a continuing discussion of issues related to the card
game Le Her. Next, in terms of ink spilt on probability, are
discussions of the ``problem of the pool,'' or Waldegrave's problem,
generalized to more than three players, and of the game Les \'Etrennes
(which may be translated as ``the gifts''). The correspondence also
contains discussions of various problems in algebra, geometry,
differential equations and infinite series.

As an aristocrat, Montmort's network included both political and
scientific connections. His letters to Bernoulli contain some
references to his political activities that sometimes kept him from
replying promptly. His brother, Nicolas R\'emond, was Chef de conseil
for Phillipe duc d'Orl\'eans, who became regent of France after his
uncle Louis XIV died in 1715 (\cite{e14}, page 599). Among the
mathematicians of the era, Montmort corresponded with Isaac Newton,
Gottfried Leibniz, Brook Taylor and Abraham De Moivre, in addition to
the Bernoullis as well as several others. As a talented amateur
mathematician, his work was well regarded by the mathematicians of his
day. He was generous to his scientific friends. He received as guests
to the Ch\^ateau de Montmort Nicolaus Bernoulli, Brook Taylor and one of
the sons of Johann Bernoulli. He also sent gifts of cases of wine and
champagne to both Newton and Taylor.

Le Her is a game of strategy and chance played with a standard deck of
fifty-two playing cards. The simplest situation is when two players
play the game, and the solution is not simply determined even in that
situation. Montmort calls the two players Pierre and Paul. Pierre deals
a card from the deck to Paul and then one to himself. Paul has the
option of switching his card for Pierre's card. Pierre can only refuse
the switch if he holds a king (the highest valued card). After Paul
makes his decision to hold or switch, Pierre now has the option to hold
whatever card he now has or to switch it with a card drawn from the
deck. However, if he draws a king, he must retain his original card.
The player with the highest card wins the pot, with ties going to the
dealer Pierre. The game can be expanded to more than two players.
Montmort [(\citeyear{e17}), pages 186--187] originally described the problem for
four players and posed the question: What are the chances of each
player relative to the order in which they make their play?

Because of the winning conditions, it is obvious that one would want to
switch low cards and keep high ones. The key is to find what to do with
the middle cards, such as seven and eight, when two players are playing
the game. In other cases, cards are clearly too low to keep or too high
to switch, being much below or above the average in a random draw.
Naturally, the threshold would be lower with more than two players.

In Part V of \emph{Essay d'analyse}, only the game with two players is
considered. Initially, Montmort and Nicolaus Bernoulli wrote back and
forth about the problem and came to the same solution. However, two of
Montmort's friends contended that this solution was incorrect. These
were an English gentleman named Waldegrave and an abbot whose abbey was
only a league and a half (about 5.8 kilometers) from Ch\^ateau de
Montmort (\cite{e18}, page 338). Montmort identified Waldegrave
only as the brother of the Lord Waldegrave who married the natural
daughter of King James II of England. Lord Waldegrave is Henry
Waldegrave, 1st Baron Waldegrave, and his wife is Henrietta FitzJames,
daughter of James II, and his mistress Arabella Churchill. The abbot is
the Abb\'e d'Orbais; Montmort also refers to him as the Abb\'e de
Monsoury. The reason for the two appellations for the abbot is that his
full name is Pierre Cuvier de Montsoury, Abb\'e d'Orbais. He has been
described as ``un prodige de bon coeur, d'urbanit\'e et de science''
(\cite{e08}). For the spelling choice between Montsoury and Monsoury,
it should be noted that Montmort often spelled his name ``Monmort.''

Two other problems were discussed extensively in the correspondence.
The first is the problem of the pool, a problem that Waldegrave
suggested to Montmort and solved himself (\cite{e18}, page 318). In
\emph{Essay d'analyse} the problem is solved for three players. It is
often called Waldegrave's problem \citep{e02}. The ``pool'' is a
way of getting three or more players to gamble against one another,
when the game put into play is for two players only. In the situation
for three players (Montmort uses the names Pierre, Paul and Jacques),
all three begin by putting an ante into the pot. Then Pierre and Paul
play a game against each other. The winner plays against Jacques and
the loser puts money into the pot. The game continues until one player
has beaten the other two in a row. That player takes the pot. The game
can be expanded to more than three players, but that situation was not
fully treated in \emph{Essay d'analyse}. The second is the problem of
solving the game Les \'Etrennes (or ``estreine,'' an alternative old
French spelling). As described by Montmort [(\citeyear{e18}), pages 406--407], this
is a strategic game between a father and his son. The father holds an
odd or even number of tokens in his hand, which his son cannot see.
When the son guesses even, he receives a gift of four \'ecus (silver
coins) if he is correct and nothing if wrong. When the son guesses odd,
he receives one \'ecu if he is correct and nothing if wrong. The
discussion of this game in the correspondence is only brought in to
enlighten Le Her whose strategic nature is in some important respects
essentially similar.

Montmort concludes the last letter (to Bernoulli) that appears in \emph
{Essay d'analyse} with a remark that Waldegrave had volunteered to take
care of getting the book printed in Paris. Montmort's letter was dated
November 15, 1713, and was written from Paris. What
is also of concern to us are the letters after this date and how these
letters relate to earlier discussions. The unpublished correspondence
begins with a letter from Montmort to Bernoulli dated January 25, 1714,
in which he says that he has sent Bernoulli two copies of the second
edition of \emph{Essay d'analyse}. Montmort was still in Paris, where
he claimed to have been for three months. He was staying at a hotel in
Rue des Bernardins, which in modern Paris is only a walk of 350 meters
to the printer, Jacques Quillau in Rue Galande. Presumably,
Waldegrave's help consisted mainly in dealing with the printer and the
proof sheets as they came off the press, thus relieving Montmort of
some tedious work.

\section{The Treatment of the Game of Le Her in \emph{Essay d'analyse}}

To understand the discussion of Le Her after 1713, it is necessary to
describe the treatment of the game as it appears in the second edition
of \emph{Essay d'analyse}. Hald [(\citeyear{e12}), pages 314--322] provides a
detailed description of the mathematical calculations involved in
assessing the game. Yet he devotes little space to elucidating the
discussions among Bernoulli, Montmort and Waldegrave, as well as the
Abb\'e d'Orbais, concerning the issues surrounding these mathematical
calculations. It is the substance of these discussions that are of
interest to us.

Hald's only comment on the discussion over Le Her concerns a comment
made by Waldegrave and Abb\'e d'Orbais to the effect that Bernoulli's
reasoning in obtaining his mathematical solution is faulty. After
pointing out their observation that Bernoulli's solution fails to
account for a player's probability of playing in a certain way, Hald
[(\citeyear{e12}), page 315] claims:

\begin{quote}
It is no wonder that Bernoulli does not \mbox{understand} the implications of
this remark,
since the writers themselves have not  grasped the full
implication of their point of view.
\end{quote}
It is indeed true that there was some confusion on Bernoulli's side
which he deftly tried to hide.

Henny [(\citeyear{e13}), page 502] comments that he is amazed to find expressed in
the letters many concepts and ideas that appear in modern game theory.
At the same time, he is
surprised to find Waldegrave defending his position so strongly against
Bernoulli who was the superior mathematician. Henny states further that
Waldegrave did not have the necessary mathematical skills to provide a
mathematical proof of his results.

As we will show in a review of the treatment of Le Her in \emph{Essay
d'analyse} and the subsequent unpublished correspondence, both Hald's
and Henny's insights fall short of the mark.\footnote{The same could be
said of others who have passed harsh judgments on Montmort and
Bernoulli. For instance, \citet{e11} argues that ``Montmort's
conclusion [that no absolute rule could be given], though obviously
correct for the limited aspect in which he viewed the problem, is
unsatisfactory to common sense, which suggests that in all
circumstances there must be, according to the degree of our knowledge,
at least one rule of conduct which shall be not less satisfactory than
any other; and this his discussion fails to provide.'' Our discussion
below will show that this assessment is misinformed.} One reason they
fall short is that they do not consider the full range of the various
events that were under discussion and their associated probabilities.
Two events are natural for a probabilist to consider. The first is the
distribution of the cards to Pierre and Paul. The second is the
randomizing device used to come up with the mixed strategy prescribing
when the players should hold and when they should switch. The
randomizing device considered in \emph{Essay d'analyse} is a bag
containing black and white counters or tokens (the old French word used
is ``jetton''). The third event that Montmort, Waldegrave and Bernoulli
consider (but not Hald or Henny) is difficult, or perhaps impossible,
to quantify. This is the possibility that Paul, say, is a poor player
and does not follow a strategy
that is mathematically optimal, or the possibility that Paul, say, is a
very good player who tries to trick Pierre into making a poor choice.
This kind of event unfolds regularly in modern poker games.

Another reason for which Hald and Henny see some confusion in the
discussions among Bernoulli, Montmort and Waldegrave is that what we
are seeing in the correspondence is the complete unfolding of a problem
from its initial statement, and discussions around it, to a complete
solution. This is different from a ``textbook'' statement of a problem
followed by a solution. In the latter case, the problem and solution
are both well laid out. In the former case, there is some grappling
with the problem until it becomes clear how to proceed.

 We begin with the correspondence in \emph{Essay}  \emph{d'analyse}
 where Le Her
is first mentioned. In Johann Bernoulli's 1710 letter to Montmort in
\emph{Essay d'analyse}, he suggests more efficient methods to reach
Montmort's conclusions for a variety of problems and in some cases
generalizes Montmort's results. There is only one reference to the
problem of Le Her, which is the second of four problems proposed in
Montmort [(\citeyear{e17}), pages 185--187]:

\begin{quote}
The second and the third [problem] seem to me amenable, but not without
much difficulty and work, that I prefer to defer to you and learn the
solution, than to work long at the expense of my ordinary occupations
that leave me scarcely any time to apply myself to other things.
\end{quote}
In his reply to this letter, which is dated November 15, 1710
(\cite{e18}, pages 303--307), Montmort makes no reference to this passage.

Nicolaus Bernoulli's first letter to Montmort, dated February 26, 1711,
makes no reference to the game Le Her. It is a note in Montmort's reply
to Nicolaus Bernoulli, dated April 10, 1711 (\cite{e18}, pages
315--323) that initiates the discussion of Montmort's second problem:

\begin{quote}
I started some time ago to work on the solution of problems that I
propose at the end of my book; I find that in Le Her, when there are
only two players left, Pierre and Paul, Paul's advantage is greater
than 1 in 85, and less than 1 in 84. This problem has difficulties of a
singular nature.
\end{quote}
In a postscript to this letter, Montmort makes an additional remark:

\begin{quote}
As there are few copies of my book left, there will soon be a new
edition. When I have decided, I will ask you permission, and your
uncle, to insert your beautiful letters which will make the principal
embellishment.
\end{quote}
It is this announcement that may have motivated Nicolaus Bernoulli to
continue his correspondence with Montmort and to send him much
interesting material. Publishing mathematical material outside a
scientific society or without a patron to cover the costs was an
expensive proposition, one that Montmort could afford. Because of the
specialized type that was used and the accompanying necessary skill of
the typesetter, the cost of a mathematical publication was well above
the norm for less technical books. Bernoulli could get his results in
print at no cost to himself.

Bernoulli responded with a long letter, dated November 10, 1711
(\cite{e18}, pages 323--337). In this letter, he announces, among
many other things, that he has also solved the two-person case for Le
Her (\cite{e18}, page 334):

\begin{quote}
I also solved the problem on Le Her in the simplest case; here is what
I found. If we suppose that each player observes the conduct that is
most advantageous to him, Paul must only hold to a card that is higher
than a seven and Pierre to one that is higher
than an eight, and we find under this supposition that the lot of
Pierre will be to that of Paul as 2697 is to 2828. Supposing that Paul
also holds to a seven, then Pierre must hold to an eight, and their
lots will still be as 2697 to 2828. Nevertheless it is more
advantageous for him not to hold to a seven than to hold to it, which
is a puzzle that I leave you to develop.
\end{quote}
This passage is carefully worded, yet it will be misinterpreted by
Montmort and Waldegrave. As we will see, a key aspect that is neglected
by Montmort and Waldegrave is the antecedent of Bernoulli's conditional
statement starting with ``If we suppose that each player observes the
conduct\ldots''

Montmort's reply, dated March 1, 1712 (\cite{e18}, pages 337--347),
highly praises Ber\-noulli's prior letter. He complains that, being in
Paris, he has had no time and peace to think on his own and, as a
consequence, the main object of his letter is to report progress made
by his two friends, the Abb\'e d'Orbais and Waldegrave, on a problem
proposed by Bernoulli, and on the problem of Le Her. On the latter,
Montmort reports that ``they dare however not submit to your
decisions'' (\cite{e18}, page 338). However, as he says in a
passage that is key to understanding the forthcoming controversy, the
Abb\'e d'Orbais also previously disagreed with Montmort:

\begin{quote}
When I worked on Le Her a few years ago, I told M. l'Abb\'e de Monsoury
what I had found, but neither my calculations nor my arguments could
convince him. He always maintained that it was impossible to determine
the lot of Pierre and Paul, because we could not determine which card
Pierre must hold to, and vice versa, which results in a circle, and
makes in his opinion the solution impossible. He added a quantity of
subtle reasonings which made me doubt a little that I had caught the
truth. That is where I~was when I proposed that you examine this
problem; my goal was to make sure from you of the goodness of my
solution, without having the trouble of recalling my ideas on this
which were completely erased.
\end{quote}
Montmort then claims that Bernoulli's solution confirms what he had
found, a decision that prompts a reply from Waldegrave objecting to
Bernoulli's solution, quoted at length in Montmort (\citeyear{e18}), pages 339--340.

According to Waldegrave and the Abb\'e d'Orbais, it is not true that
Paul must hold only to an eight and Pierre to a nine. Rather, that Paul
should be indifferent to hold to a
seven or to switch, and that Pierre should be indifferent to hold to an
eight or to switch. Waldegrave wrote the following to Montmort
(\cite{e18}, page 339):

\begin{quote}
We argue that it is indifferent to Paul to switch or hold with a seven,
and to Pierre to switch or hold with an eight. To prove this, I~must
first explain their lot in all cases. That of Paul having a seven, is
$\frac{780}{50\times51}$ when he switches, and when he holds on to it
his lot is $\frac{720}{50\times51}$ if Pierre holds on to an eight, and
$\frac{816}{50\times51}$ if Pierre switches with an eight. The lot of
Pierre having an eight is $\frac{150}{23\times50}$ if he holds on to
it, and $\frac{210}{23\times50}$ if he switches in the case that Paul
only holds on to a seven; and $\frac{350}{27\times50}$ by holding on to
it, and $\frac{314}{27\times50}$ by switching in the case that Paul
holds on to a seven, so here they are. The lots of Paul $\frac{780\
\mathrm{or}\ 720\ \mathrm{or}\ 816}{50\times51}$, those of Pierre $\frac
{150\ \mathrm{or}\ 210}{23\times50}$ or $\frac{350\ \mathrm{or}\
314}{27\times50}$.
\end{quote}
Based on the numbers he obtains, Waldegrave observes that ``720 being
more below 780 than 816 is above, it appears that Paul must have a
reason to switch with 7'' (\cite{e18}, page 339). The differences,
$780-720$ and $816-780$, are in the ratio $60\dvtx 36$, or $5\dvtx 3$, a ratio
which later enters the discussion.

In the rest of his argument, Waldegrave talks of a weight instead of a
reason. He first lets the weight that leads Paul to switch be $A$, and
the weight that leads Pierre to switch be $B$. And he argues that the
same weights lead Paul and Pierre to both strategies. $A$ leads Paul to
switch with 7 and, as a consequence, it also leads Pierre to switch his
8; but what leads Pierre to switch his 8 must lead Paul to hold with 7.
So, $A$ leads Paul to both switch with a 7 and hold on to it. The same
goes for Pierre. Therefore, ``it is false that Paul must only hold on
to an 8, and Pierre to a 9,'' which was Bernoulli's claimed solution.
The word ``probability'' comes up only once in this discussion, in the
conclusion
of the excerpt from Waldegrave's letter to Montmort. Waldegrave writes
(\cite{e18}, page 340):

\begin{quote}
Apparently Mr. Bernoulli was simply looking at the fractions that
express the different lots of Pierre and Paul, without paying attention
to the probability of what the other will do.
\end{quote}
Montmort leaves the discussion there without further comment.

\begin{table}[t]
\tablewidth=240pt
\tabcolsep=0pt
\caption{Probabilities that Paul wins depending on the strategies of play}\label{table1}
\begin{tabular*}{240pt}{@{\extracolsep{\fill}}lcc@{}}
\hline
\\[-16.5pt]
\multirow{2}{*}{\backslashbox{\textbf{Paul}}{\textbf{Pierre}}}
&\textbf{Switch the 8} &\textbf{Hold the 8}
\\
&\textbf{(and under)}&\textbf{(and over)}
\\[-2.5pt]
\hline
Switch the 7 &\multirow{2}{*}{$ \frac{2828}{5525}$}&\multirow{2}{*}{$
\frac{2838}{5525}$}
\\
\quad(and under) &&
\\
Hold the 7&\multirow{2}{*}{$ \frac{2834}{5525}$}&\multirow{2}{*}{$ \frac
{2828}{5525}$}
\\
\quad(and over) &&
\\
\hline
\end{tabular*}
\end{table}

Upon receiving Montmort's letter, Bernoulli agrees with these figures,
saying that ``the lots they found for Pierre and Paul are very right''
(\cite{e18}, page 348). And yet, when Bernoulli proposes his
solution, and when Montmort eventually publishes a table of
probabilities as an appendix to \emph{Essay d'analyse} (\cite{e18},
page 413), the numbers are different. The Bernoulli--Montmort
probabilities are shown in Table~\ref{table1}, which appears in Hald
(\citeyear{e12}), page 318. None of the parties in this debate actually explain
their calculations. Waldegrave's probabilities are justified in
Todhunter (\citeyear{e21}), pages 107--110; the Bernoulli--Montmort probabilities
are in Hald (\citeyear{e12}), pages 315--318. The difference in the probabilities
is that Waldegrave's probabilities are conditional on Paul having a
seven in his hand and the Bernoulli--Montmort probabilities are the
marginal probabilities for all cards that Paul may hold.

In a letter dated June 2, 1712, Bernoulli replies to Waldegrave's
argument by accusing him of committing a fallacy. He argues that if we
suppose that $A$ leads Paul to switch with a seven, and so leads Pierre
to switch with an eight (if Pierre knows Paul switches with seven),
then it also leads Paul to hold on to a seven. Therefore, $A$ both leads
Paul to switch with a seven and to hold on to a seven. His conclusion
is that (\cite{e18}, page 348):

\begin{quote}
we are supposing two contradictory things at the same time; that is,
that Paul knows and ignores at the same time what Pierre will do, and
Pierre what Paul will do.
\end{quote}
Bernoulli explains that if we do not commit this fallacy regarding what
Paul and
Pierre know about the other's intent, we are led to reasoning in a
circle, which shows that Waldegrave's argument cannot show anything.
This argument is peculiar, and seems to suggest that Bernoulli does not
understand Waldegrave's point. It might, however, be simply a
misinterpretation of Waldegrave's argument, for it is expressed in
terms of weight rather than in terms of probability. The word
``weight'' or ``poids'' in French offers more opportunity for
misinterpretation. Moreover, Bernoulli admits having written his letter
hastily, as he was preparing for a long trip through the Netherlands
and England. As a result of this travel, some subsequent letters are
delayed, and the arguments they contain do not follow the chronological
order of when the letters were written.

A letter to Bernoulli, dated September 5, 1712 (\cite{e18}, pages
361--370), announces that Waldegrave and the Abb\'e d'Orbais have seen
Ber\-noulli's reply in which he accuses them of committing a fallacy.
Montmort includes a note from the Abb\'e d'Orbais in which he claims
that Waldegrave has written a beautiful and precise reply to
Bernoulli's objection; the rebuttal, however, is not included. In this
note, the Abb\'e d'Orbais also enjoins Montmort to take a side in this
dispute between them. This suggests that, even if Montmort thanked
Bernoulli for his solution, which he claimed agreed with his own,
Montmort has not yet made up his mind as to whether Bernoulli really
solved the problem.

The next letter concerning Le Her is from Bernoulli to Montmort, dated
December 30, 1712 (\cite{e18}, pages 375--394). Adding important
pieces to the puzzle, it contains a three-page discussion of Le Her
(Bernoulli mentions having just received the June 2 letter, since it
was sent from Switzerland to Holland, then to England, and finally back
to Switzerland). Bernoulli insists that, despite Waldegrave's
arguments, Paul does not do as well by abiding to the maxim of holding
to a seven, than that of switching with a seven. Bernoulli then says
(\cite{e18}, page 376):

\begin{quote}
If it were impossible to decide this problem, Paul having a seven would
not know what to do; and to rid himself [from deciding], he would
subject himself to chance, for example, he would put in a bag an equal
number of white tokens and black tokens, with the intent of holding to
a seven if he draws a white one, \& to switch with a seven if he draws
a black one; because if he put an unequal number he would be lead more
to one party than to the other, which is against the assumption. Pierre
with an eight would do the same thing to see whether he must switch or not.
\end{quote}
This comment introduces with clarity the idea of chance by ``the way of
tokens'' (as they will say later). What Bernoulli says here seems to
confirm that, at first, when he accused Waldegrave of committing a
fallacy, he did not interpret Waldegrave's weights as probabilities.
Nonetheless, he suggests that the only probability allocation
compatible with the supposed state of ignorance of the players is that
each player chooses a strategy with probability $\frac{1}{2}$. Under
these choices, he computes the lot of Paul (which is then $\frac
{774}{51\times50}$) and concludes that it would be a bad thing for Paul
to randomize in this way, since he could guarantee himself a lot of
$\frac{780}{51\times50}$. Therefore, Bernoulli concludes Paul must
always switch with a seven. As Bernoulli says (\cite{e18}, page
376), ``it is better to make the choice where we risk less.'' He then
explains the reasoning that he had left out of his hastily written
letter from June 2. In contemporary terms, he calculated the
unconditional probability of winning under each pure strategy profile
(without assuming that any card has been dealt yet). He displays a
refined version of the reasoning that led to accusing Waldegrave of a
fallacy, yet it does not do full justice to Waldegrave's idea.

Eight months later, on August 20, 1713, Montmort [(\citeyear{e18}), pages 395--400]
finally replies to Bernoulli, complaining that he has, despite his
philosophical inclinations, been involved in political activities, and
so he did not have the leisure for intellectual work. Thus, his letter
only contains scientific news. There is only one brief mention of Le
Her; he tells Bernoulli that, despite his last effort to provide a
thorough and precise argument, Waldegrave and the Abb\'e d'Orbais are
still unconvinced by his claimed solution. Shortly after, in a letter
dated September 9, 1713, Bernoulli also asks Montmort to explain his
own views on the dispute. Montmort obliges him in his letter dated
November 15, 1713.
This is the last letter published in the second edition of \emph{Essay
d'analyse} (\cite{e18}, pages 403--413). The letter also contains an
excerpt of a letter from Waldegrave and a table of the lots of Paul and
Pierre for the four crucial combinations of strategies, which are
summarized in Table~\ref{table1}.

Here, then, is Montmort's understanding of the situation. To begin
with, he agrees with Bernoulli that it is not indifferent to Paul to
switch or hold with a seven, and to Pierre to switch or hold with an
eight, because of Bernoulli's calculations of the unequal chances for
each strategy. (This shows that Bernoulli and Montmort use
``indifferent'' in the sense of having the same probability of winning.
For Waldegrave and d'Orbais, however, ``indifferent'' seems to mean,
perhaps more awkwardly, that no strategy dominates the other in
probability.) This being said, Montmort nonetheless disagrees with
Bernoulli that this establishes the strategy as a maxim, that is, as a
rule of conduct that must be obeyed invariably to obtain the best
results. Rather, he thinks that it is impossible to establish such a
maxim (\cite{e18}, page 403):

\begin{quote}
[T]he solution of the problem is impossible, that is, we cannot
prescribe to Paul the conduct that he must adopt when he has a seven,
and to Pierre the conduct he must adopt when he has an eight.
\end{quote}
He grants that, if one is to choose a fixed and determined maxim, then
switching on seven, for Paul, will be better than any other, yet Paul
can hope to make his lot better.

Why, then, would a solution be impossible? Would the solution not be
the optimum that one can reach in Paul's hope of making his lot better?
Montmort claims that, whereas he used to think that the use of black
and white tokens to randomize strategies could avoid the ``circle,'' he
does not think that anymore. He gives a general formula to find the
probability of winning with a certain probability allocation for what
we call a mixed strategy:
\[
\frac{2828ac+2834bc+2838ad+2828bd}{13\cdot17\cdot25(a+b+c+d)} ,
\]
where $a$ is Paul's probability of switching with seven, $b$ is Paul's
probability of holding the seven, $c$ is Pierre's probability of
switching with an eight, and $d$ is
Pierre's probability of holding on to an eight. But how should the
probabilities be chosen? Montmort claims that any argument will only
inform us of what Paul must do conditionally to what Pierre does and
vice versa, which leads us into a circle once again. He concludes that
Bernoulli's arguments to show that a circle does not occur are wrong,
and instead formulates this thesis (\cite{e18}, page 404):

\begin{quote}
[W]e must suppose that both players are equally subtle, and that they
will choose their conduct only based on their knowledge of the conduct
of the other player. However, since there is here no fixed point, the
maxim of a player depends on the yet unknown maxim of the other, so
that if we establish one, we draw from this supposition a contradiction
that shows that we must not have established it.
\end{quote}
Montmort also disagrees with Bernoulli that, under pain of
contradiction, if we are to use white and black tokens to randomize, we
must use an equal number of tokens. Instead, he thinks that the
probability of winning calculated for the fixed and determined maxims
shows that Paul must switch more often with a seven than hold on to it.
Yet, he maintains (\cite{e18}, page 405):

\begin{quote}
But how much more often must he switch rather than hold, and in
particular what he must do (here and now) is the principal question:
the calculation does not teach us anything about that, and I take this
decision to be impossible.
\end{quote}
Thus, Montmort believes, it seems, that there is no optimal probability
allocation.

But he has another reason for believing that the solution of the game
is impossible. He has in mind the game Les \'Etrennes (\cite{e18},
pages 406--407). Montmort also believes that it is impossible to
prescribe any strategy of play in Les \'Etrennes because the players
might always try, and indeed good players will try, to deceive other
players into thinking that they will play something they are not
playing, thus trying to outsmart each other (``jouer au plus fin'' is
the phrase used in French).

As he was finishing his letter, Montmort received one from Waldegrave
and quoted extensively from it to Bernoulli. \emph{Essay d'analyse}
essentially concludes with Waldegrave's letter. Waldegrave refers to a
formula, which is not included by Montmort; it presumably is the
formula displayed above. He explains that, if $a=3$ and $b=5$ (so that
the probability of Paul switching with a seven is $0.625$), then the
lot of Pierre is
going to be $\frac{2831}{5525}+\frac{3}{4\cdot5525}$ no matter what $c$
and $d$ are. This shows that $\frac{2831}{5525}+\frac{3}{4\cdot5525}$
is Paul's minimum lot. He can only adopt another conduct in the hope of
making his lot better. This shows, he claims, that both Bernoulli and
(formerly) himself were wrong to claim that the lots of Paul was to
that of Pierre as 2828 is to 2697; if both players play in the most
advantageous way, Paul's lot is $\frac{2831}{5525}+\frac{3}{4\cdot
5525}$. Waldegrave is convinced that this is something that both
Bernoulli and Montmort will agree to, now that it is agreed that one
can use a randomized strategy. He also explains that, if Pierre uses
$c=5$ and $d=3$, then $\frac{2831}{5525}+\frac{3}{4\cdot5525}$ will
also be Paul's maximum lot.

Waldegrave also asserts that it is impossible to establish a maxim; he
grants, however, that it is impossible for him to show this with the
same level of evidence. This is often taken incorrectly as evidence of
a lack of Waldegrave's mathematical abilities. Waldegrave is instead
referring to the situation in which players may try to outsmart each
other. Waldegrave agrees that if Paul does not use $a=3$ and $b=5$,
then it is possible for Paul to do better than $\frac{2831}{5525}+\frac
{3}{4\cdot5525}$ provided that Pierre does not play in the best way. On
the other hand, it would be worse if Pierre plays correctly.
Furthermore, Waldegrave remarks (\cite{e18}, page 411):

\begin{quote}
What means are there to discover the ratio of the probability
that Pierre will play correctly to the probability that he will not?
This appears to me to be absolutely impossible, and thus leads us into
a circle.
\end{quote}
As with Montmort, his main concern is that it is always possible for
the players to try to outsmart each other (``jouer au plus fin'').

\section{Issues Arising from the Published Correspondence}

Examining the detailed arguments provided by  Montmort, Bernoulli and
Waldegrave reveals a picture that contrasts with the judgment that they
were essentially confused on the fundamental concepts and methods
required to solve a strategic game such as Le Her. In fact, we maintain
that they understood most of the aspects of the problem with clarity.
There are, however, a number of important outstanding issues left
unresolved in the correspondence on Le Her as it appears in \emph{Essay
d'analyse}. Let us review them briefly.

It is true that the letters reveal a certain type of misunderstanding;
however, it is not conceptual confusion, but rather mutual
misinterpretation due to using terms differently. An instance of this
is whether it is indifferent to Paul to switch or hold to a seven. On
the one hand, both Montmort and Bernoulli claim that it is not
indifferent to Paul because the chances of winning are not identical.
On the other hand, Waldegrave claims that it is indifferent, and the
reason for that seems to be that neither pure strategy dominates the
other in probability.

Another instance of this is the disagreement they appear to have on the
existence of a circularity in the analysis of the game. Montmort and
Waldegrave assert that there is a vicious circle that prevents one from
establishing a maxim; the circle they discuss, however, is really a
regression \emph{ad infinitum}, that is, to establish a maxim, we
always need to go one step further in the ``$A$ must know what $B$
does'' loop (Bernoulli agrees with this point). However, Bernoulli
claims that there is a circle in Waldegrave's argument, in the sense
that either his argument is contradictory or a \emph{petitio principii}
(but Bernoulli is not considering randomizing strategies at this
point). Again, they are only contradicting each other in the wording,
not in the idea.

Finally, a third instance is that Montmort and Waldegrave claim that
the solution of the game is impossible, whereas Bernoulli does not.
Here again, they disagree on what it means to ``solve'' the game Le
Her. Bernoulli claims that the solution is the strategy that guarantees
the best minimal gain---what we would call a minimax solution---and
that as such there is a solution. However, despite understanding this
``solution concept,'' Montmort and Waldegrave refuse to affirm that it
``solves'' the game, since there are situations in which it might not
be the best rule to follow, namely, if a player is weak and can be
taken advantage of. Clearly, the concept of solution they have in mind
differs from
the minimax concept of solution. This latter concept, in addition to
the probability of gain with a pure strategy and the probability
allocation required to form mixed strategies, requires that we know the
probability that a player will play an inferior strategy. But, they
assert, this cannot be analyzed by calculations, so the game cannot be solved.

This being said, there are a number of things that are said that
suggest a certain level of confusion at a conceptual level. The two
most important are these. First, Bernoulli appears to have some
difficulty with the relation between the knowledge of the players and
the probabilities involved in mixing strategies. His circularity
objection to Waldegrave is awkward and somewhat mystifying. Moreover,
his argument that, if we allow randomized strategies with black and
white tokens, it must be because neither player knows what the other
player will do, and that as a result the only acceptable probability
allocation of $\frac{1}{2}$ is problematic. This kind of mistaken
argument has been repeated over the centuries by some of the greatest
minds in probability, statistics and game theory. Second, Montmort
understands very well the idea of randomizing strategies, but he
nonetheless claims that there is no optimal probability allocation that
can be calculated. This claim, however, was made before consulting
Waldegrave's letter in which he reveals the optimal probability.

\section{Discussion of the Game of Le Her After~1713}

Referring to a letter from Bernoulli to Montmort dated February 20,
1714, \citet{e13}, in his treatment of Le Her, mentions only that
Bernoulli accepted Waldegrave's solution to the problem. However,
Bernoulli had other things to say about Le Her in that same letter.
Henny also refers to a letter of January 9, 1715, from Waldegrave to Bernoulli
in which Waldegrave seemingly admits to Bernoulli that he does not have
the mathematical skills to actually prove his results. What Henny
leaves out is that the letter was written in reply to a detailed
criticism of the solutions to Le Her that Bernoulli had sent earlier to
Montmort.

After some personal news and apologies for not writing sooner, in his
letter of February 20, 1714, to \mbox{Montmort}, Bernoulli initially thanks
Montmort for correcting, editing and making clearer his letters that
Montmort had printed in \emph{Essay d'analyse}. Then follows the
discussion of Le Her that \citet{e13} only briefly mentions.
Initially, Bernoulli suggests that the controversy is essentially over:

\begin{quote}
Concerning Le Her, I seem to have foreseen that in the end we would all
be right. However, I congratulate Mr. de Waldegrave who has the final
decision on this question, and I willfully grant him the honor of
closing this affair\ldots
\end{quote}
Despite this, Bernoulli still claims that he disagrees on a few minor
points, and these point directly to the outstanding issues we mentioned
above. The main concern is the relation between ``establishing a
maxim'' and solving the problem of Le Her posed by Montmort in his
book. Bernoulli states:

\begin{quote}
One can establish a maxim and propose a rule to conduct one's game,
without following it all the time. We sometimes play badly on purpose,
to deceive the opponent, and that is what cannot be decided in such
questions, when one should make a mistake on purpose.
\end{quote}
This point was raised before by Montmort and Waldegrave, but they do
not consider that such a play would be necessarily a mistake. Whether
or not this kind of play is a mistake, we saw that from the same
consideration, Montmort and Waldegrave conclude that solving the
problem is impossible. However, Bernoulli now phrases things more carefully:

\begin{quote}
Mr. de Waldegrave wrongs me on p. 410 by claiming that I once said that
the lot of Paul is to that of Pierre as $2828\dvtx 2697$. If you carefully
read my letter from Oct. 10, 1711, you will find that I did not say it
absolute and without restriction. I beg you to consider those words:
once we have determined or rather supposed what are the cards to which
the players will hold, etc. And the following words. You will see that
I there supposed that the players want to hold to a fixed and
determined card, and indeed I
had not thought about the way of tokens, which, as Mr. de Waldegrave
said, is not among the ordinary rules of the game.
\end{quote}
Bernoulli essentially says that he was misinterpreted and that he only
computed the best odds of winning with a pure strategy, not that he
established what a player should do in an actual game. Moreover, if we
grant his supposition, then he has found the most advantageous maxim.
After this correction, Bernoulli thinks the discussion is over, saying,
``We are thus all agreeing, and we have made peace; \emph{canamus
receptui} [sing retreat].''

In his response to Bernoulli, dated March 24, 1714, Montmort concurs by
writing, ``I am quite pleased that we are all together by and large
agreeing.'' In this letter Montmort claims that he disagreed with
Bernoulli on some aspect of the corrected interpretation of his
position, but he leaves it to a later letter to explain. However, in
his next letter to Bernoulli, November 21, 1714, Montmort does little
to clarify. He says, ``if it is ever permitted to say to two persons
maintaining contradictory claims that they are both right, it is
assuredly at this occasion in our dispute.'' Montmort emphasizes that
what he seeks is the correct advice that should be given to the
players, but the discussion does not go much further.

On August 15, 1714, Montmort sent a letter to Bernoulli containing a
two-page ``supplement'' that reignites the debate. He makes six points.
First, he claims that telling Paul always to switch with a seven is bad
advice, since his minimum lot is then 2828. Second, that it would be
better advice to tell him to do whatever he pleases with a seven, so
that he can look at both options indifferently. Third, we cannot say
that this would be the best advice either, for knowing that, Pierre
would switch with an eight, in which case Paul should certainly have
held on to a seven. This leads to a vicious circle. Fourth, if we admit
the way of tokens, the best advice that he knows is to tell Paul to
have the ratio $3\dvtx 5$ for switching with a seven. But even then, he does
not think that we can demonstrate that it is the best advice. Fifth, he
claims that it is impossible at this game to determine the lot of Paul,
because one cannot determine what manner of playing is the most
advantageous to each player, even when we admit a randomized strategy.
This point makes explicit for the first time Montmort's (and presumably
Waldegrave's) idea that you can only claim that you have found the lot
of a player (which is what Montmort's
problem in \emph{Essay d'analyse} demanded) if we can determine what is
the best way to play. Moreover, determining the best way to play
demands knowing more than the optimal token ratio for the randomized
strategy. He adds that, of course, some methods of playing are better
than others, as informed by the chances that have previously been
calculated. He concludes, sixth, that he would not know what advice to
give Paul if he had to. This letter sharpens the debate, in that it
makes explicit the connection between ``solving'' a game and giving
advice for play in actual situations.

In a long letter to Montmort dated August 28, 1714, along with a
``supplement'' dated November 1, 1714, Bernoulli replies to Montmort
point by point. He asks Montmort a question that is meant to dismiss
his argument:

\begin{quote}
If, admitting the way of tokens, the option of 3 to 5 for Paul to
switch with a seven is the best you know, why do you want to give Paul
other advice in article 6? It suffices for Paul to follow the best
maxim that he could know. It is not enough to claim that there is still
a circle despite my reasons, one must fight my reasons.
\end{quote}
And he continues: ``It is not impossible at this game to determine the
lot of Paul.'' To counter Montmort's previous argument, he once again
insists that either Paul knows what Pierre will do, in which case his
maxim is clear, or he does not, in which case Paul should use the
probability $\frac{1}{2}$ in the randomized strategy to determine what
to do. As he admits, this is the exact same position he had at the
beginning of the discussion, supported by the exact same argument.
Thus, it seems that Bernoulli has missed the point Montmort made
explicit in his August 15 letter.

It is at this point that Waldegrave reenters the debate at Montmort's
request. In a letter dated January 9, 1715, Waldegrave reiterates the
six points that Montmort had laid out for Bernoulli in his letter of
August 15. For each of the six points, Waldegrave's arguments are
longer and more detailed than what Montmort had previously given.

It is not until March 22, 1715, that Montmort replies to Bernoulli on
this dispute. It is part of a very long letter that also contains the
main topic for their further correspondence, infinite series. In this
letter, Montmort writes once again about his views. They are the same
as what we have seen already. However, Montmort stresses that a lot of
what remains under discussion is based on inconsistent terminology and
\mbox{misinterpretation}. In essence, he believes that the outstanding
disagreements are only
apparent contradictions. Nonetheless, he introduces one more element to
clearly \mbox{articulate} his view. He distinguishes between the advice that
he would put in print, or give to Paul publicly, and the advice he
would give so that only Paul hears it. Montmort claims that, for the
former, he would choose the mixed strategy with $a=3$ and $b=5$, since
it is the one that demonstrably brings about the lesser prejudice.
However, he explains that, in practice, if Paul is playing against an
ordinary player and not a mathematician, he would quietly give
different advice that could allow Paul to take advantage of his
opponent's weakness. As he explains, the objective of this sort of
analysis is not only to provide a maxim to otherwise ignorant players,
but also to warn them about the potential advantages of using finesse.
However, this latter part is not possible to establish, and it is in
this sense that there is no possible solution to this problem.

The next letter, sent by Bernoulli to Montmort on May 4, 1715,
disregards Montmort's nuance. To begin, Bernoulli ``is forced to admit
that he does not precisely know on what point [they] contradict each
other.'' Nonetheless, Bernoulli explains that, in his view, the
distinction between public and private advice, the possibility of using
finesse, or something similar, does not alter the fact that $a=3$ and
$b=5$ is the best solution, and that it determines the lot of Paul (so
that not only is the game solvable, but it is indeed solved).

Despite Bernoulli's explanation, Montmort's next letter, dated June 8,
1715, once again reiterates that ``you have badly solved the proposed
question, or you have not solved it at all.'' He makes explicit what he
takes the proposed question to be:

\begin{quote}
The question is and has always been to know whether we can establish
the lots and as a result the advantage of playing first under the
supposition not that Pierre and Paul follow this or that maxim (this
would have no utility, no difficulty), but that both of them having the
same skills, each follow the conduct that is the most advantageous.
\end{quote}
Montmort then says that this dispute is beginning to bore him. He
considers that furthering it will not make them learn anything new and
that in the end the dispute must be about some other thing.

Our presentation of the correspondence makes it clear that they are
using different concepts of solution; Bernoulli's in essence is the
concept of the minimax solution, whereas
Montmort's further depends on the probability of imperfect play (i.e.,
on the skill level of the players).

Around this time, Montmort's interest shifts from probability and its
applications to infinite series. In fact, most of the remaining
correspondence with  Bernoulli turns to that topic. At the same time,
Montmort began an extensive correspondence with Brook Taylor, also
mainly on infinite series (St. John's College Library, Cambridge,
TaylorB/E4). Although the dispute with Bernoulli seems to have petered
out, Montmort was not yet done with it. In a letter dated July 4, 1716,
Montmort asked Taylor to examine his dispute with Bernoulli about Le
Her and to express his opinion on who was right. He referred Taylor
only to the correspondence that appears in \emph{Essay d'analyse}.
Taylor apparently wrote back but with the wrong impression about what
Montmort wanted. Montmort replied to Taylor on August 4, 1716, that he
did not want any new research into the problem but only to examine, at
his leisure, which of Bernoulli or Montmort was right. In a letter to
Taylor dated November 10, 1717, Montmort thanked Taylor for his opinion
on the dispute and concluded the letter by saying that Waldegrave would
write him about Le Her as well as another game. Unfortunately, neither
Taylor's reply expressing his opinion nor Waldegrave's letter to Taylor
are extant.

\section{The Problem of the Pool and Other Probability Problems}

Compared to the discussion of Le Her, the remaining discussion in the
post-1713 correspondence with regard to probability problems is
relatively minor. For example, after the remarks on Le Her that
Bernoulli made in his letter of February 20, 1714, to Montmort,
Bernoulli comments that he thinks there is an error in Montmort's
solution to a problem related to the jeu du petit palet in \emph{Essay
d'analyse}. He asks Montmort to check his solution. The problem appears
to be Probl\`eme IV in Montmort (\citeyear{e18}), page 254. The jeu du petit
palet is a game in which players toss coins or flat stones (the
``palets'') toward a target set on the ground or a table. The player
with the most coins or stones on the target wins. The English
equivalent game is called chuck-farthing or chuck-penny.

What takes up much of the discussion, other than Le Her, is news about
Abraham De Moivre's work. De Moivre corresponded with both Montmort and
Bernoulli until
about 1715 when he ceased corresponding with either of them. Prior to
this discussion, Bernoulli had sent De Moivre a general solution to the
problem of the pool on December 30, 1713 (\cite{e03}, pages 106--107).

A report on De Moivre's activities in probability takes up part of a
letter from Bernoulli to Montmort dated April 4, 1714. Bernoulli
mentions that De Moivre has sent him a long letter with reports of new
solutions that will appear in a much expanded version of his treatise
\emph{De mensura sortis} (De Moivre, \citeyear{M11}). De Moivre's new work, which
was entitled \emph{The Doctrine of Chances}, did not appear until 1718
(De Moivre, \citeyear{M18}). No details are given to Montmort other than that De
Moivre has made inroads in three areas. First, De Moivre used his own
method for the solution of the problem of the pool to generalize it to
more than three players. Second, he developed a new kind of algebra to
solve probability problems. Finally, Bernoulli reports that De Moivre
considered that nearly all problems in probability can be reduced to
series summations. Not only did De Moivre report that he had
generalized the problem of the pool, but he also sent Bernoulli his
solution to the problem. At the time of his writing to Montmort,
Bernoulli had not read the solution and did not pass the solution on to
Montmort. The new algebra is probably the one that De Moivre developed
for finding probabilities of compound events. See, for example, Hald
[(\citeyear{e12}), pages 336--338] for a modern discussion of this topic. This part
of the letter ends with what might be interpreted as a nasty comment
about De Moivre:

\begin{quote}
I will share here in confidence what he wrote to me concerning you.
Here is what he told me about your comments that I had sent him. `I
cannot stop myself etc. Our Society etc. I just received etc. kind
[regards].' After the letter I find written there these words: `in a
sense,' that made me laugh.
\end{quote}
It is difficult to know what exactly Bernoulli is saying here. It
appears that he sent De Moivre Montmort's severe criticism of \emph{De
mensura sortis} that Montmort published in \emph{Essay d'analyse}
(\cite{e18}, pages 363--369).

Later that month, Montmort reported back to  Bernoulli that he received
a very polite and fair letter from De Moivre in which De Moivre
announced that he had found a new solution to the problem of the
duration of play. See Bellhouse [(\citeyear{e03}), pages 111--114] for a discussion
of the publication of this solution. De Moivre sent reports about more
of his results in probability to Montmort and Montmort sent on a pr\'
ecis of these results to Bernoulli in a letter dated August 15, 1714.
Many of the results that Montmort mentions found their way into \emph
{The Doctrine of Chances}, including what is called Woodcock's problem
discussed in Bellhouse [(\citeyear{e03}), pages 125--126].

\begin{figure*}

\includegraphics{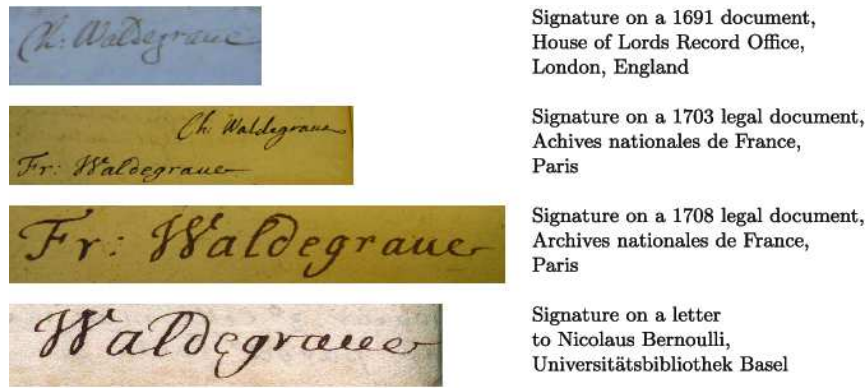}

  \caption{Waldegrave signatures from various sources.}\label{table2}
\end{figure*}

On August 28, 1714, Bernoulli finally wrote to Montmort enclosing a
copy of De Moivre's general solution to the problem of the pool. In the
letter, Bernoulli asks Montmort to tell him what he thinks of the
solution. He further states that it appears that De Moivre is using
Bernoulli's approach to the solution for three and four players that
appears in \emph{Essay d'analyse} (\cite{e18}, pages 380--387). At
the same time he is using an analytical approach rather than infinite
series (De Moivre actually used a recursive method for his general
solution). Montmort replied on March 22, 1715, that he agrees with
Bernoulli's assessment. On returning from a trip to England, Montmort
reported to Bernoulli in a letter dated June 8, 1715, that one of
Bernoulli's solutions to the problem of the pool had just been printed
in the \emph{Philosophical Transactions} (Bernoulli, \citeyear{e07}). Bernoulli had sent De Moivre
two solutions; De Moivre claimed he had found an error in the first solution.

\section{Waldegrave Identified}

Many in the past have tried unsuccessfully to identify the Waldegrave
who solved the problem of Le Her and who suggested the problem of the
pool, often called Waldegrave's problem. \citet{e02} reviewed
these attempts at identification and narrowed the field down to
Charles, Edward or Francis Waldegrave, the three brothers of Henry
Waldegrave, 1st Baron Waldegrave. Bellhouse argued for Charles
Waldegrave, but in view of new information his choice was incorrect.
Key to the proper identification is that several
Waldegraves---siblings, cousins and at least one uncle of
Henry---followed King James II into exile in France after James was
deposed in 1688.

The proper identification of the Waldegrave of interest may be found in
legal papers in the Archive nationales de France in conjunction with a
letter from Waldegrave to Nicolaus Bernoulli; the letter to Bernoulli
is signed only ``Waldegrave'' and is the only known letter in
Waldegrave's hand that is extant (Universit\"atsbibliothek Basel L Ia
22, Nr. 261). Other Waldegrave signatures to compare to the one on
Bernoulli's letter can be found on various legal documents, two in
France (Archives nationales de France MC/ET/XVII/486 and 514) and one
in England (House of Lords Record Office HL/PO/JO/10/1/439/481).
See Figure~\ref{table2}.
From the signatures, it is obvious that Francis is the Waldegrave of
interest. From these records, it is also apparent that Charles
Waldegrave handled the family's affairs in England while Francis
Waldegrave took charge of them in France.

What little is known of the life of Francis Waldegrave comes mostly
from Montmort's correspondence with Brook Taylor and Nicolaus
Bernoulli. Montmort reported to Taylor one of Waldegrave's political
activities. Waldegrave was planning to take part in the Jacobite
uprising in England in 1715. He was to be part of an invasion force led
by the son of James II, James Stuart. The uprising in England fizzled
out, James Stuart remained in France and Waldegrave fell ill just prior
to the time when the planned invasion was to occur. Montmort called
Waldegrave's illness apoplexy; it was probably a stroke. From time to
time, Montmort commented to Taylor and Bernoulli about Waldegrave's
illness, recovery and setbacks. At one point, for a cure or a rest,
Waldegrave took the waters at a spa in France. He also spent time at
Montmort's chateau. Though ill, he was alive in France in 1719 when
Montmort died so that the flow of information to Bernoulli and Taylor
about Waldegrave stopped. Presumably, Waldegrave died in France.

How Waldegrave obtained his mathematical training is unknown. In
whatever way he was educated, he was an adept amateur mathematician.
This is contrary to Henny's interpretation of Waldegrave's skills. For
example, Henny [(\citeyear{e13}), page 502] claims that Waldegrave did not have the
mathematical skills to work out a general method of calculation in Le
Her. On the contrary, there is a hint of the fairly high level of Waldegrave's
mathematical abilities in a letter from Montmort to Bernoulli dated
March 24, 1714. There Montmort says that he is getting
Waldegrave to read L'H\^opital's (\citeyear{e15}) calculus book \emph{Analyse des
infiniment petits} and that Waldegrave has a natural aptitude for
mathematics.

At the time that Montmort was sending \emph{Essay d'analyse} to his
publisher, Francis Waldegrave was living in Rue Princesse near \'Eglise
Saint-Sulpice in Paris. In modern
Paris, it is a $1.1$ kilometer walk to Montmort's publisher in Rue
Galande. In Section~\ref{sec1} it was mentioned that Montmort was staying only
$350$ meters from his publisher. This was not the only time that
Montmort enlisted a colleague to take on some of the tedious parts of
getting results to print. After Montmort sent Brook Taylor a number of
theorems about infinite series, they decided that the results should be
published in the Royal Society's \emph{Philosophical Transactions}
\citep{e16}. In a letter dated June 15, 1717, Montmort gave Taylor
complete editorial control over the paper that included having Taylor
translate the results from French into Latin (St. John's College
Library, Cambridge TaylorB/E4). Taylor replied August 9, 1717, saying
that he had made many changes and corrections to the paper (St. John's
College Library, Cambridge, TaylorB/E5).

\section{Discussion and Conclusions}

The unpublished letters between Bernoulli and  Montmort reveal a much
more complex story than either \citet{e13} or \citet{e12} have
described. The entire group---Bernoulli, Montmort and Wal\-degrave---were
for the most part clear about the issues at the conceptual level. In
the end it came down to a disagreement about what it meant to solve a
problem. Further, Henny recognized many modern game theory concepts,
but we show that the group's understanding of the modern notions is
deeper than what Henny realized.

Apart from the technical and conceptual aspects of Le Her and other
probability problems, we also get a glimpse into the social side of a
rich amateur mathematician at work. Montmort was a good mathematician,
but mathematics was his hobby and at times he did not have time to
pursue his hobby. There is a bit of quid pro quo in his relationships
with Bernoulli, Taylor and Waldegrave. Montmort acquires some status
through his connections to artists, philosophers and scientists. He can
impose on his scientific friends to do some of the more menial work for
him in getting his research to print. On the other side, his scientific
friends enjoy his hospitality, his gifts and the benefits of any
political and scientific connections that he may have.

Traditionally, the mixed strategy solution with $a=3$ and $b=5$ for Le
Her has been attributed to Waldegrave. It certainly appears to be the
correct attribution based on the correspondence in the second edition
of \emph{Essay d'analyse}. However, in the long
letter from Montmort to Bernoulli dated March 22, 1715, that covers
discussions of Le Her, De Moivre and other topics, Montmort appears to
claim priority of solution. As part of the discussion of Le Her, he
says, ``although I first found the determination of the numbers $a$ and
$b$, $c$ and $d$\ldots'' Montmort's suggestion of priority could have
come about as a result of a conversation between Waldegrave and
Montmort, with Waldegrave putting pen to paper. This illustrates
Fasolt's (\citeyear{e10}) claims about the limits of history. Our data from the
past is what has been written, not what has been spoken. Further, we
can never know the tone behind what was written, such as Bernoulli's
apparently nasty comments to Montmort about De Moivre in his letter of
April 14, 1714. Instead of coming up in conservation, Montmort may be
claiming priority because he found the general formula in $a$, $b$, $c$
and $d$; the numbers were only a special case. Or it could be something
else. Like Le Her itself, depending on how the problem is approached,
the assignment of priority is a problem with no solution.

\section*{Acknowledgments}

We would like to thank Dr. Fritz Nagel of Universit\"at Basel for
giving us access to the correspondence between Nicolaus Bernoulli and
Pierre R\'emond de Montmort. We also thank Kathryn McKee and Jonathan
Harrison of St. John's College Library, Cambridge, for providing us
with copies of the correspondence between Brook Taylor and Montmort.

The originals of the letters of Bernoulli, Montmort and Waldgrave are
in Universit\"atsbibliothek Basel. The letters from Montmort to
Bernoulli are catalogued Handschriften L Ia 22:2 Nr.187--206 and from
Bernoulli to Montmort are L Ia 21:2 Bl.209--275. The letter from
Bernoulli to Waldegrave is catalogued L Ia 21:2 Bl.229v--232r and the
letter from Waldegrave to Bernoulli is L Ia 22, Nr. 261. When
referencing these letters, we have to do so by the date, writer and
recipient, rather than the catalog numbers.


%

\end{document}